# People are not coins: Morally distinct types of predictions necessitate different fairness constraints

Eleonora Viganò
University of Zurich
eleonora.vigano@ibme.uzh.ch

Corinna Hertweck
University of Zurich and Zurich University of Applied Sciences
corinna.hertweck@zhaw.ch

Christoph Heitz
Zurich University of Applied Sciences
christoph.heitz@zhaw.ch

Michele Loi
Politecnico di Milano
michele.loi@polimi.it

## ABSTRACT

In a recent paper [1], Brian Hedden has argued that most of the group fairness constraints discussed in the machine learning literature are not necessary conditions for the fairness of predictions, and hence that there are no genuine fairness metrics. This is proven by discussing a special case of a fair prediction. In our paper, we show that Hedden's argument does not hold for the most common kind of predictions used in data science, which are about people and based on data from similar people; we call these "human-group-based practices." We argue that there is a morally salient distinction between human-group-based practices and those that are based on data of only one person, which we call "human-individual-based practices." Thus, what may be a necessary condition for the fairness of human-group-based practices may not be a necessary condition for the fairness of human-individual-based practices, on which Hedden's argument is based. Accordingly, the group fairness metrics discussed in the machine learning literature may still be relevant for most applications of prediction-based decision making.

## 1 INTRODUCTION

With the increasing use of automated decision-making algorithms in high-stakes domains, such as hiring, lending and education, fairness has to be a principal component in the design of these algorithms. In the past years, the algorithmic fairness literature has proposed a variety of fairness metrics, which may help to gauge the fairness of decision-making systems [2], [3]. While fulfilling these metrics is typically intuitively appealing, the literature has also shown that they are often mutually incompatible [4], [5]. This gives rise to questions such as which metrics should be evaluated and whether fulfilling any of these metrics is a necessary condition for fairness. Brian Hedden's recent paper "On Statistical Criteria of Algorithmic Fairness" [1] provides good reasons to question whether some of the most widely considered statistical fairness constraints (also called group fairness metrics or criteria) – collectively labelled as the set F – are necessary conditions for the fairness of probability estimates. In this paper, we show that the conclusion made in Hedden's paper is not applicable to most automated decision-making systems analyzed by the algorithmic fairness literature. While the analyzed fairness criteria are indeed not necessary conditions for fairness for all automated decision-making systems, they are still relevant and might even be necessary for most of today's decision-making systems.

Typically, decisions made under uncertainty are based on the assessment of the probabilities of the possible outcomes. There are two ways of computing probabilities, which depend on the data that we use. When we want to determine the probability that a coin lands heads up, we toss it several times and then divide the number of times it landed heads up by the total number of throws. When deciding whether a defendant has a low risk of recidivism, we might look at the behavior of other individuals that had a case similar to the defendant and that are similar to the latter with respect to certain characteristics. The coin and the recidivism cases are different because in the former case, we can toss the coin as many times as we want, whereas in the latter case the defendant's probability of recidivism cannot be assessed only on the basis of the data about the considered defendant, but requires data from similar defendants because we cannot release the defendant more than one time.

On the basis of the distinction between the two ways of computing probability, we distinguish two kinds of probabilities: individual-based and group-based probabilities. In this contribution, we will focus on predictions and decisions regarding humans, thus we will deal only with human-individual-based probabilities (for simplicity, h-individual probability) and human-group-based probabilities (for simplicity, h-group probability). An h-individual probability is determined with data of the person whom the prediction is about; an h-group probability is determined with data of other people beyond the person whom the prediction is about. These two types of h-probabilities correspond to the two types of probability-based decision-making practices: an h-individual practice is one in which we make a decision about a person by employing the h-individual probabilities attributed to that person; an h-group practice is one in which we take a decision about a person by employing the h-group probabilities attributed to that person.

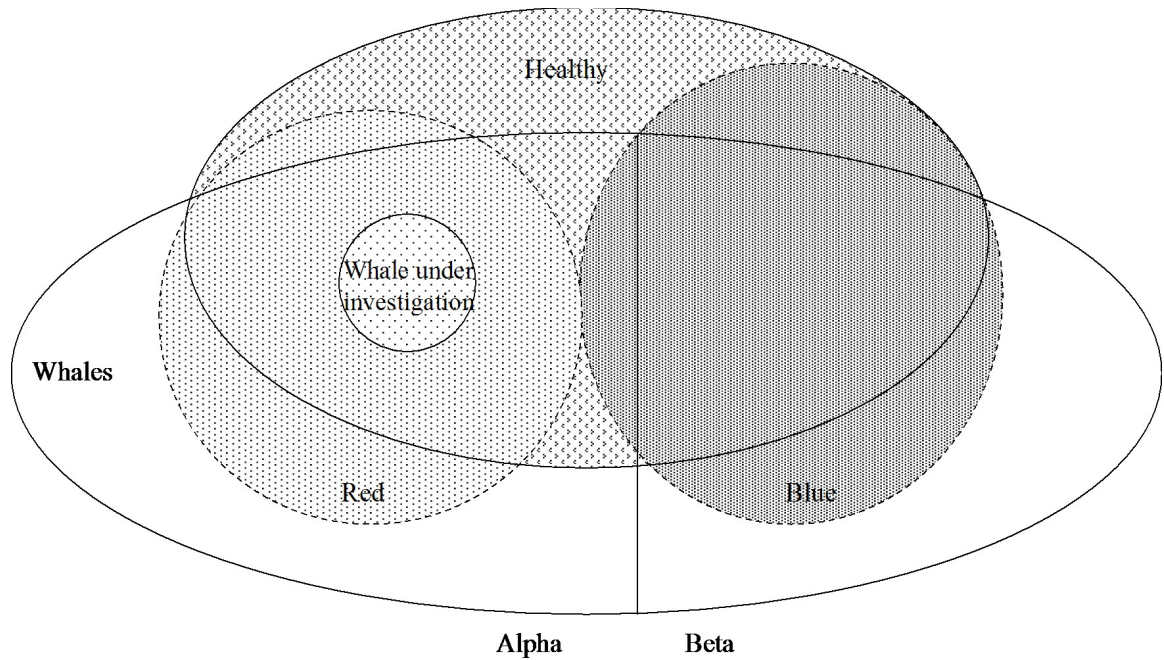

Figure 1: The whale set is dissected into two sub-sets (alpha- and beta-whales) by the straight line. These are the two types of whales. Being blue is necessary for beta-whales to be healthy, but it is not necessary for alpha-whales to be healthy.

Hedden does not indicate what kind of probability estimates – and thus practices, as defined above – he considers. In this paper, first, we argue that Hedden's argument is valid for h-individual practices, but it fails for h-group practices. Then, we claim that there is a moral difference between h-individual probabilities and h-group probabilities and thus between the two kinds of h-practices based on them. We show that, for h-group practices being considered fair, all things considered, additional conditions with respect to misclassification harm may be necessary, and that at least one of the statistical fairness constraints does provide such a condition. This proves that, for h-group practices, Hedden's argument is not valid and the statistical fairness constraints might play a decisive role in determining the fairness of a prediction. Note that we do not contend that the two h-probabilities are distinct at the epistemic level (thus, we do not take a stance about the strength of a probabilistic belief), nor at the ontological level (thus we do not contend that the two h-probabilities are metaphysically distinct), as our argument does not require such distinctions.

In the following, we use an example to illustrate the logic of our argument against Hedden's contention that most of the statistical fairness constraints are not necessary conditions for the fairness of predictions. Suppose that someone claims that a whale is only healthy if it is blue. Blueness is thus claimed to be a necessary condition of whale health. Suppose we now learn that there are two kinds of whales: alpha-whales and beta-whales. As it turns out, alpha-whales can also be healthy if they have other colors including red. The existence of whales that are red and healthy shows that blueness is not a necessary condition for the health of alpha-wales. Since it is not a necessary condition for the health of alpha-whales, it is not a necessary condition for the health of whales in general. This finding, however, does not disprove that blueness is a necessary condition for the health of beta-whales – it still could be, since we have not gained any new knowledge about beta-whales that would prove otherwise (see Fig. 1).

We are able to make a logical statement about whales in general – namely, being blue is not necessary for them to be healthy. This is certainly something useful to know, especially when confronting someone who seems unreasonably confident that, since some given whale is not blue (without investigating further if it is of the alpha or beta kind), it must necessarily be sick. However, this should not lead anyone to disregard blueness as necessary for the health of beta-whales, which is especially relevant if beta-whales are much more common than alpha-whales. Hedden's argument functions in a similar way in that it only proves that most fairness metrics $f_n$ in the set F are not necessary for h-individual practices; as we will show, he conceives an h-individual practice that is fair and fulfills only one fairness metric, which is calibration. However, Hedden does not prove that most fairness metrics are not necessary for h-group practices, which we claim are much more common and relevant in practice (see Table 1).

Table 1: Analogy between types of whales and types of probability-based practices

| Whales | Probability-based practices |
| --- | --- |
| Alpha whales | H-individual practices |
| Beta whales | H-group practices |
| Healthy | Fair |
| Blue | Fairness metrics $f_n$ in F that are necessary conditions for the fairness of h-group practices (e.g., equalized odds) |
| Red | Calibration |
| Whale under investigation | Practice identified by Hedden |

To show this, we first summarize Hedden's argument in Section 2. Then, in Section 3, we describe the logical structure of our argument. In Section 4, we explain the distinction between h-individual and h-group probabilities in more detail. In Section 5, we describe h-individual and h-group practices and anchors the theoretical discussion of decisions based on probabilities in actual practice of data science. In Section 6, we justify the claim that there is a moral distinction between h-individual and h-group practices. We claim that h-group practices could be seen as pro tanto morally wrong because they violate principle M, which is a person's pro tanto right of being treated as an individual, while h-individual practices are not vulnerable to this moral objection. Section 7 is a reply to the main objections against principle M and the moral distinction in h-practices. Section 8 concludes the paper.

## 2 HEDDEN'S ARGUMENT: MOST GROUP FAIRNESS METRICS ARE UNNECESSARY CONDITIONS FOR FAIRNESS

Hedden makes the argument that many group fairness metrics are unnecessary conditions for fairness. For this, he constructs a case in which people are assigned coins which can either land heads up or tails up. This coin can only be thrown once [1]. If it lands heads up, the person is deemed a "heads person." Each coin has a label which shows its probability of landing heads up. A predictor assigns risk scores to people meant to indicate their risk of being a heads person. This risk score is then used to predict the person as either being a heads person (risk score above 0.5) or a tails person (below 0.5). What the predictor does to assess the risk score is to simply echo the label of the person's coin. This is an intuitively and perfectly fair approach. Hedden then brings in the idea of measuring fairness with respect to socio-demographic groups: people are sent to two different rooms, which represent the socio-demographic groups with respect to which we want to measure fairness. In one room, the coin labels are close to 0 and 1, in the other room the labels are closer to 0.5. This allows Hedden to construct a case with equal average coin labels.

Due to the different distributions of labels, many fairness metrics (10 out of the 11 he analyzed) are violated in Hedden's perfectly fair coin example. The only metric that is fulfilled is calibration. In this case, calibration requires that for a given risk score, the ratio of people who have been assigned this risk score and in fact turn out to be heads people is roughly equal to this risk score. For the risk score 0.8, for example, 80% of all the people who have been assigned this risk score should be heads people. This is clearly fulfilled if this is tested on enough people since a coin label of 0.8 means that it will land heads up in 80% of the cases. Hedden thus concludes that the fairness metrics in F cannot possibly be genuinely necessary for fairness, with the exception of calibration, which is fulfilled in Hedden's perfectly fair coin example.[1]

## 3 OUTLINE OF THE ARGUMENT AGAINST HEDDEN'S ARGUMENT

The argument in this paper has the following form:

- P1. One can distinguish between two practices for determining[2] probabilities to be assigned to persons: those that use data of other people (h-group probabilities) and those that only use data of the person to whom the probability is assigned (i.e., h-individual probabilities).
- P2. Hedden's coin example is an instance of using an h-individual probability. He shows that, for any $f_n$ in F, $f_n$ is not necessary for the fairness of h-individual predictions. By way of implication, $f_n$ is not necessary for fairness in general.
- P3. Many practices in data science involve h-group probabilities. This includes the COMPAS example that Hedden treats as paradigmatic.
- P4. Practices involving h-group probabilities differ from practices involving h-individual probabilities from the moral point of view.
- P5. The fact that some $f_n$ is not a necessary condition for the fairness of probability-based practices in general does not logically exclude that $f_n$ may be a necessary condition for the fairness of h-group-based practices.

Conclusion: Hedden has not offered an argument for data scientists using h-group probabilities to disregard F as a necessary condition for the fairness of their algorithms.

---

[1] Note that Hedden does not contend that calibration is necessary for the fairness of any practice. He rather shows that the practice that he conceives is fair and satisfies calibration but no other fairness metrics.
[2] Our usage of the term "determining" here refers to all processes that lead us to assign probabilities. This includes testing, which is necessary for the assignments to be considered scientifically justified. When using machine learning to assign these probabilities, the term refers to the entire machine learning pipeline, i.e., data collection, data cleaning, training, testing, etc.

The argument for premise P5 has already been provided in the introduction and has been made more comprehensible by the whale analogy. The next sections provide support for the other premises: Section 4 will argue for P1 and P2, Section 5 for P3, and Section 6 for P4.

## 4 INDIVIDUAL-BASED- AND GROUP-BASED-PROBABILITIES

P1. One can distinguish between two practices for determining probabilities to be assigned to persons: those that use data of other people, and those that only use data of the person to whom the probability is assigned.

P2. Hedden's coin example is an instance of using an h-individual probability. He shows that, for any $f_n$ in F, $f_n$ is not necessary for the fairness of h-individual predictions. By way of implication, $f_n$ is not necessary for fairness in general.

Probabilities are used to deal with situations where there is uncertainty in the outcome. In the case of algorithmic decision making, where a decision maker wants to take a decision with respect to a human individual, we are dealing with the uncertainty of current or future features of persons (such as whether a person commits a crime). In Hedden's coin example, he associates a probability with each coin, "indicating its bias, or its objective chance of landing heads" [1, p. 219].

While bracketing all philosophical considerations about the meaning of the concept "probability," we assume that there is something like an objective chance $x$ of landing heads which is given for each individual coin, as Hedden suggests. Under this assumption, we can actually associate this probability $x_i$ with each individual $i$, in the sense of a feature of this individual. Actually, many prediction-based decision algorithms make exactly this kind of association, and the associated probability is taken as an input for the decision, as Hedden suggests for his "perfectly fair and unbiased predictive algorithm" [1, p. 219].

A frequent problem in practice, however, is that it is often not possible to determine such a probability based on the data of a single individual, or it is extremely costly to do this. Take, for example, measuring the probability that a 50-year-old individual will die of cancer within the next 10 years, which would be a typical prediction task for a machine learning algorithm. This probability depends on the complete genetic predispositions of this individual, as well as on everything that happens in their life. Since there was never a person on earth with the same genotype and the same life experiences, it is impossible to determine the probability by considering only that individual, even if we had access to all the information about them.

In the case of simple systems such as coins, the standard way of determining probabilities for a specific outcome to be predicted – such as landing heads – is to repeat some experiment many times: we throw the coin in a well-defined manner under well-defined environmental conditions many times and count how often it lands heads or tails, respectively. Under the assumption that the coin does not change its nature during these observations and afterwards (stationarity), we can then determine the probability of landing heads up to an arbitrary precision. After this is done, we can treat this probability as a feature of the coin, and label it accordingly. It is easy to see that this is indeed a feature of this specific coin, as only this coin is used to determine its value. However, for most events that are relevant in data-based prediction of individuals, there is just not enough history to observe enough repetitions. We cannot observe enough lifetime of an individual for determining their probability of committing a crime in the next month to any reasonable accuracy, letting alone the obvious violation of the stationarity assumption. Furthermore, as the aforementioned example shows, in some cases, repetitions over a lifetime are often just not possible.[3]

In cases in which we cannot observe enough repetitions, an alternative approach may be chosen. We combine observations of different individuals for determining the searched-for probability. For example, we may observe a sample of 1000 50-year-old people over a period of 10 years and record how many of them die of cancer during this period. The percentage of deceased people is then interpreted as a probability, and we assign all individuals of the group this probability. In fact, this probability is extremely useful because it allows us to make predictions: if we take a new group of 50-year-old persons, selected according to the same selection procedure as the first group, we can predict the expectation value of how many of them will die, together with confidence intervals (under some additional assumptions). For the coin example, the alternative approach means that instead of repeating a coin throw experiment 1000 times with the same coin, we take 1000 identical coins and throw each once, recording on which side the coin lands. Statistical theory tells us that taking the number of observed heads and dividing them by 1000 yields a non-biased estimate of the true probability of each of these coins – assuming that all coins do have the same probability (which is meant by "identical coins").[4] A similar approach is used in the training phase of practically any data-driven prediction algorithms: observations of events of numerically distinct but similar individuals are analyzed together, and the resulting estimated probability is then assigned to each of those individuals.[5] This is done in all cases where repetitions are not possible, but also in many other cases, where collecting data of different individuals over a short period is just easier, cheaper, or more convenient than observing single individuals over a long period (and then assuming that the recorded past can be extrapolated to predict the future).

Thus, for determining the searched-for probability, we can either apply a practice that involves only data (observations) of the individual under consideration, namely the h-individual probabilities, or a practice that also involves data of other individuals, namely the h-group probabilities. More generally, also when not referring to humans, we may speak of individual-based probabilities and group-based probabilities. However, the general distinction is not morally salient the way – we shall argue – the human distinction is. In the case of deciding what to do with our coins, for example, the distinction between individual-based and group-based probabilities is not morally important.

---

[3] We do, however, rely on individual histories when making informed guesses about individuals we know. For example, we may think it probable that a friend who has very often been late to appointments in the past will be late in the future. This can be described as an intuitive probabilistic belief, but is not formalized as an application of statistical methods.

[4] The distinction we draw does not map neatly into any crucial distinction in the philosophy of probability. However, Eells uses similar examples to illustrate the difference between *token* and *type probabilistic causality* [6]. Token causation claims typically refer to particular individuals, places, and times (e.g., the smoking of this cigarette increased the probability of cancer by 0.1%) or – with reference to Hedden's example – the probability of this coin to land heads up caused the procedure to select the coin owner for a prize. Type causation claims typically refer to types of events, e.g., "smoking causes cancer" or being a coin of this type (i.e., coin with p>0.5 to land head) causes the procedure to select the coin owner for a prize. Arguably, token causality claims should be supported by evidence about the single individual involved in causation, while type causality claims can be supported by evidence collected about the entire class of similar causes.

[5] In practice, this is a bit more complicated. Prediction models assume some kind of smoothness or functional dependency and then interpolate the training data.

It is now easy to show that in Hedden's coins example, the risk scores assigned to people are h-individual probabilities. Hedden starts with "coins of varying biases" [1, p. 219], making clear that the bias is a generic property of the coin itself. Accordingly, the coin can be labelled with this probability. Then he states: "each individual in the population is randomly assigned a coin" [1, p. 219]; this means that the probability assigned to each person (which is the probability that their coin lands heads) does not depend in any way on anything related to other people. In such a case, Hedden's statement that this assignment is perfectly fair and unbiased is true. Yet, as we will show in the next section, the use of h-individual probabilities is not very common in algorithmic practices.

## 5 PROBABILITY IN MACHINE LEARNING

P3. Many practices in data science involve h-group probabilities. This includes the COMPAS example that Hedden considers to be paradigmatic.

Let us now turn our focus to machine learning and discuss what kinds of probabilities we encounter in this field. There are different subfields of machine learning, but what is relevant for our paper is the usage of machine learning to make predictions based on input data. This is what the subfield of supervised machine learning is concerned with. For this, a data set consisting of input features $x$ and target variables $y$ exists. This could be, for example, data about individuals who have been granted a loan by a given bank in the past. Each row in this data, i.e., each accepted loan application, comes with a label $y$, which is the target for the prediction task. If the goal in the banking example is predicting whether a given applicant will repay their loan on time, $y$ would be a label for just whether the individual repaid their loan on time. The goal of supervised machine learning is to find a mapping from the input features $x$ to the target variables $y$.

To do this, a large enough data set needs to be provided. This data is typically split into two parts: a training set and a testing set. An algorithm "trains" on the training data set, meaning it infers rules about the relationship of $x$ and $y$ from the training data. The algorithm defines what these rules might look like. In logistic regression, for example, a sigmoid curve is fit to the data. The "rules" would then correspond to the weight assigned to the different input features. Other machine learning algorithms might fit other curves or models to the training data. The algorithm keeps adjusting these rules until the predictions appear to work well for the training data. Finally, the rules it found are tested on the test set to ensure generalizability.

The data set of input features $x$ and target variables $y$ can consist both of data from a single individual or many different individuals. In the case that the data set consists of data from a single individual, we speak of an h-individual practice. In the case that the data set consists of data from more than one individual, we speak of an h-group practice. The advantage of h-group practices is that the data collection process is easier and faster. Moreover, the prediction model of h-group practices is more generalizable. With enough data, such a model could, for example, be applied to any individual – as opposed to individualized models which are expected to only be valid for the single individual whose data they were trained on. More importantly, however, for a lot of cases, it is simply impossible to train a model on the data of only one individual. For this reason, h-group practices are extremely common in data science. This includes the COMPAS example that Hedden specifically refers to. In the case of COMPAS, data from thousands of defendants who were admitted for bail were collected along with the target label $y$ of whether they would have been arrested within the next two years. This target variable $y$ is supposed to represent recidivism. Here, it would clearly be not only inefficient, but also impossible to build a machine learning model based on only a single individual's data. The developers of the tool therefore trained their model based on data of many defendants. The case thus represents an instance of an h-group practice.

Hedden states that he is "not claiming that the case of people, coins, and rooms is realistic or completely analogous to cases like COMPAS. Of course, it is not" [1, p. 223]. He then goes on to explain several reasons why the coin example is not completely analogous to COMPAS, but "idealizes away from these complications" [1, p. 224]. What he does not appear to consider though is the question of whether the probability labels of coins are comparable to the probability outputs of COMPAS and what influence this comparison has on the claims he makes. Our argument is that the coins' probabilities and the defendants' assigned probabilities are different in such a way that the claims Hedden proves for the coin example are merely applicable for cases of h-individual practices, but not for h-group practices, such as COMPAS, which are very common in machine learning.

## 6 THE MORAL DIFFERENCE BETWEEN H-INDIVIDUAL PRACTICES AND H-GROUP PRACTICES

P4. Practices involving h-group probabilities differ from practices involving h-individual probabilities from the moral point of view.

Let us consider the case of a car insurance company that offers its customers personalized prices based on their attributes (e.g., driving behavior) and that can use two different algorithms for assigning the personalized insurance premium to the person A: model 1 and model 2. Model 1 represents an h-individual practice as it has only been trained on A's past driving behavior (e.g., involvement in accidents that were A's fault). Model 2 represents an h-group practice as it has been trained on the data of many individuals, some of whom are similar to A under some characteristics (e.g., gender, age), in case there is a statistically sound association between accidents and, for example, a combination of such characteristics (e.g., people in their sixties have the fewest accidents).[6] It seems that model 2 is

---

[6] It may be objected that when the target variable is causally affected by group membership – e.g., men might cause more accidents than women – predicting this target variable and using it in decision-making is prima facie unfair. Yet in this our contribution, we focus on a specific argument for prima facie unfairness, namely, the two ways of computing probability (either by using data from other individual or data relative to

morally problematic because it uses information from other individuals to make a decision about A: the latter is judged on the basis of characteristics and information of which A is not responsible (e.g., it is not A's fault if people of A's age drive dangerously and A should not pay for their behaviors). By contrast, model 1 (i.e., the h-individual practice) uses only information regarding A, which is about what A did (e.g., complying with traffic regulations) or some other characteristics of A (e.g., being a prudent driver). This prima facie moral distinction between h-group and h-individual practices is based on the intuition that predicting a person's behavior based on data of other persons (h-group probabilities) is wrong. This intuition was found in Binns and colleagues' qualitative study on people's perceptions of algorithmic decision-making [7]. While some subjects agreed on using information about an individual's past behavior to make predictions about their future behavior, they objected about using information about other individuals to make the same predictions [7, pp. 377-8].

Because of the prima facie distinction between predicting one's behavior using other persons' data and predicting one's behavior only using data of the considered person, we cannot simply assume that practices generating or employing h-individual and h-group probabilities are morally analogous. We rather need to delve into the intuition that h-individual and h-group practices are morally different due to their different ways of estimating probabilities about people. Such an intuition has been defended in moral philosophy and legal studies by means of a moral requirement or duty, which we call principle M. The latter demands that people should be treated as individuals [8], [9].

The understanding of principle M requires a clarification of what we mean by "individual" when this term is attributed to people. In principle M, a person's individuality concerns them being that particular person and not another one, namely them being a unique individual. After having clarified the meaning of "individual," when the latter is applied to people, we need to specify what the right attitude towards this individual is, namely, how to treat a person properly, where "properly" means in a way that is appropriate to what this individuality is. We shall use the expression "respect for individuality" here, to indicate this mode of treatment that is compatible with treating people as individuals. It is not essential for this argument to determine what this expression amounts to exactly. However, it seems plausible that, whatever it means, it cannot imply treating the individual based on the assumption that they will act similarly to how other people who are similar to them in some respects have acted. This is because doing so would intuitively amount to treating the individual as a specimen of a more general category or group.

In the legal field, principle M is discussed with regards to the value of statistical evidence for legal proof in courtrooms. The main issue in this field is the weight of statistical evidence for proving a defendant guilty [10]. Several legal scholars contended that statistical evidence cannot ground verdicts because using statistical evidence to form a judgment about a defendant violates principle M [11]–[13].[7]

In moral philosophy, principle M has been investigated in the context of statistical discrimination [8], [9], [16, ch. 11], racial profiling[8] – which is an instance of statistical discrimination – and stereotyping [17], [18]. According to Eidelson, principle M is a moral requirement that demands treating the individual as partly the results of their previous choices with which they determined their life [9, p. 216]. Lippert-Rasmussen conceived principle M as the demand to make accurate judgments about a person, using all the relevant information reasonably available to one [8, p. 54]. Beeghly contended that principle M in the forms provided by Eidelson and Lippert-Rasmussen can be considered a moral obligation "sometimes or always" [18, p. 708]. Moss considered profiling as an epistemic failure that has a moral implication [10, p. 221], which is the violation of an interpretation of principle M: the *rule of consideration* [10, p. 223]. This is a moral norm stating that, in forming beliefs about a person, one should consider the possibility that this person, as an individual, can be an exception to statistical generalization deriving from statistical evidence [10, p. 221]. According to Basu, beliefs about a person based on statistical evidence violate the moral demand of relating to others as people, which is her understanding of principle M [19, p. 928]. Similarly, in Blum, principle M is understood as a fundamental form of acknowledgment of persons, whose violation is a moral fault [17, pp. 272–273].

The scholars that defended principle M as a moral requirement did not argue that it is an absolute moral principle. It is considered a pro tanto principle that can be overridden by other moral obligations that are relevant in the context of action [8, p. 57], [10, p. 224], [19, p. 927], disregarded in case of limited or no individual evidence (i.e., information about the individual person) [8, p. 54], [9, p. 224], and not required in fleeting interactions with people [17, p. 282]. As principle M is a pro tanto principle, its violation is morally permissible – even required – when the benefits of holding a belief grounded on statistical evidence overcome the harm deriving from treating somebody just as a member of a group, or when holding this belief keeps society safe. For instance, an emergency room doctor who has to decide the best treatment for a person is morally required to violate principle M if the use of statistical evidence based on the person's race for choosing the treatment increases the chance to save this person's life [10, p. 224]. Moreover, the scarce availability of information about the individual person makes the violation of principle M permissible. When getting information about an individual is costly or not possible, one is all-things-considered morally justified to base one's belief on that individual on statistical evidence derived from other people's data [8, p. 54], [9, p.224].

Now that we have identified and explained principle M, we can see in detail in which sense it traces a moral difference between h-individual and h-group practices. H-individual practices are based on h-individual probabilities, which are estimated only on the basis of what the person did (e.g., being late or punctual, thrifty or

---

the individual under consideration); such ways do not depend on the causal influence of group membership on target variables.

[7] Other scholars defended the opposing view, i.e., using statistical evidence is not necessarily a failure to treat the defendant as an individual, where principle M is conceived as respecting people's autonomy. See [14], [15].
[8] In profiling, one forms or acts on beliefs based on statistical evidence about a trait or behavior of a person. In racial profiling, one forms or acts on beliefs based on statistical evidence about a person's race.

spendthrift) or other characteristics (e.g., financial independence, stability of employment, having dependent children) that do not require comparisons with the outcomes of other people to reasonably inform a guess about the future outcomes of that person. Thus, h-individual practices treat a person as an individual. H-group practices are based on h-group probabilities, which are founded on the attribution of the person to a specific group sharing the same or similar aspects with that person (e.g., being a woman, a bank clerk, a member of the same age group, punctual). Thus, h-group practices treat the person as an instance of a more general type, not as an individual. In the above-mentioned case of car insurance, model 2 violates principle M as it does not treat A as an individual but rather as a specimen of a group of individuals that are considered similar to A. In contrast, model 1 treats person A as an individual because it relies only on information about A, without comparisons with any group. Certainly, even model 1 uses statistical information, for instance it makes generalizations about A from A's past behavior, but it never considers A as an instance of a class.

In conclusion, principle M requires that our judgments about people are constrained by the concept of person as a unique being. Since it is a pro tanto principle, it is not a prohibition to use statistical evidence tout court, nor an obligation to eschew generalization. Such a theoretical position thus provides support to two propositions of our approach to h-individual and h-group practices: (1) = (M): it is pro tanto wrong not to treat people as individuals; (2) making decisions about people based on h-group probabilities is pro tanto wrong because it violates principle M.

## 7 THE MAIN OBJECTIONS AGAINST PRINCIPLE M AND THE MORAL DISTINCTION IN H-PRACTICES

Some scholars in moral philosophy did not consider principle M as a moral requirement and its violation as morally wrong [20]–[22]. According to Arneson [20, p. 787] and Levin [21, p. 23], principle M is not morally required as using statistical evidence to form a belief about a person is not (even pro tanto) morally wrong. According to Schauer [22, p. 19], principle M may not even be initially plausible. He contended that the difference between treating people individually and based on statistical evidence is not clear. This is because treating people on the basis of individualized evidence requires assuming some generalizations that ground the individualized treatment on statistical evidence [22, pp. 101, 103, 172].

Our reply is that we do not contend that one that treats people as individuals does not engage in statistical evidence. We rather contend that there are two forms of statistical knowledge. The first is the inductive generalization on an individual that is formed by observing the behavior of individuals similar to the latter (h-group probability), such as in the case in which we throw a thousand coins once. The second form of statistical knowledge is the inductive generalization on an individual that is formed by observing the behavior of that individual throughout time (h-individual probability), such as when we throw one coin a thousand times. In this second form, we do not need to consider other individuals than the one under consideration. Thus, *contra* Schauer, it is conceivable that we use statistical generalization that does not violate principle M. This is also the position defended by Miller, in the case of predictions about a candidate's future performance in a job [23]. Miller contends that a candidate's desert to get the job cannot depend on the statistical information about the group they belong to. Yet if the statistical information is about that particular individual, the use of that piece of information does not violate principle M.[9]

The lack of agreement among scholars on the validity and bindingness of principle M may be considered a weak point of our moral distinction between h-practices as the latter is based on principle M. We reply that such a distinction is not put into question by this objection because we did not argue for an ultimate moral difference. We rather point to a prima facie difference, which has been found in people's perceptions about the morality of algorithmic decision-making [7]. To our knowledge, no scholars in moral philosophy have provided a rejection of the distinction between h-group and h-individual practices. Therefore, principle M remains a plausible hypothesis explaining the reason why h-group practices are pro tanto wrong.

Finally, it may be objected that if the wrongness of h-group-practices is due to unfairness, then we should not employ them. For we know already that h-group-practices are unfair, *qua* h-group-practices; hence, we ought not to engage with them. Finding the necessary conditions for the fairness of predictions, $f_n$, that apply only to h-group-practices, would be redundant, namely it would be one explanation too many. This objection is based on the contrast strategy, which assumes, as explained by Kagan, "that if a factor has genuine moral relevance, then for any pair of cases, where the given factor varies while others are held constant, the cases in that pair will differ in moral status" [24, p. 12]. This objection, however, commits the additive fallacy, namely it assumes that in all, or at least most cases, if $f_n$ has genuine moral relevance, it must make a difference in moral status [24]. Instead, $f_n$ may have genuine moral relevance and only make a difference in moral status in some cases, but not all.

Suppose, for argument's sake, that judging individuals based on the color of their skin is always pro tanto wrong, meaning, wrong unless an overriding moral reason for doing that exists. Now, consider a case in which judging individuals based on the color of their skin (e.g., using ethnic identity information in the search of a terrorist) is necessary to prevent a severe wrong to be committed such as a nuclear bomb to be detonated [25]. While the use of skin color information can be wrong pro tanto, the action may be morally permissible all things considered, even obligatory, because the utilitarian value of the action may override the reason that makes it pro tanto wrong.

Let us turn to a similar case involving an algorithm using h-group probabilities: a prediction P involves using h-group probabilities as the basis of a decision. Given principle M, this is pro tanto wrong. Suppose now that P enables us to prevent a more serious moral wrong and using h-group probabilities raises the probabilities of success from 60% to 80%. A utilitarian approach may contribute to a justification of h-group probabilities being fair, all things considered, in this case. By contrast, let us suppose for the sake of the argument that drastically improving the probability of preventing a more serious wrong is not sufficient to make P fair; in order for P to be fair, all things considered, the misclassification harm imposed on individuals of different groups must also be proportionate, whatever that means exactly [25]. This is not a prima facie implausible view, especially when the technical means to make it proportionate (e.g., some algorithm guaranteeing that specific statistical criteria be satisfied) exist, and if experts could agree on a definition of what "proportionate" means.[10] If, as required by [25], the fairness of P depends on the fact that P brings about proportionate misclassification harm, it is morally plausible that one of the statistical fairness constraint $f_n$, e.g., equalized odds, can be a necessary condition for the fairness of using h-group probabilities. The statistical criterion $f_n$ makes a moral difference here because P is pro tanto unfair and can be fair all things considered by virtue of the utilitarian argument, but only if $f_n$ is satisfied. Satisfying $f_n$ could, indeed, be what makes the misclassification harm imposed on different groups proportionate.[11] We do not claim that experts know already what that $f_n$ is, but Hedden's discovery pertaining to an instance of h-individual probability does not offer experts any reason to exclude any of the putative fairness constraints included in F.

---

[9]Miller adds that other moral principles, though, would prohibit the employer to reject a female candidate because she could get pregnant in the future even when the employer bases their judgment on statistical information about the female candidate as an individual. For instance, considerations of social justice would require the employer to bear part of the costs of raising a kid by hiring the woman and supporting her during the career break [23, note 24, p. 309].

Therefore, in the case of h-individual practices, as there is one possible fair use of h-individual probabilities, even if it violates $f_n$, $f_n$ is not a necessary condition for the fairness of all predictions. In the case of h-group practices, they are fair only if proportionate attention to different groups is ensured, which can be specificized by a given statistical criterion $f_n$ that may be part of F.

## 8 CONCLUSION

In our paper, we distinguished h-individual and h-group probabilities and h-individual and h-group practices. We argue that Hedden's contention that most fairness metrics in F are not necessary conditions for the fairness of algorithmic practices holds in case of h-individual practices but not in case of h-group practices. As the latter are the most common predictions in data science, while h-individual practices are the exception rather than the rule, we claim that Hedden's contention may not be too significant in practice. For all we know, in the majority of data practices, some fairness metrics could still be worth treating as moral constraints and there could be important moral arguments against their violation.

Furthermore, we claimed that there is a prima facie moral difference between computing a probability based on data of other persons deemed to be similar (h-group probabilities), and only using data of the considered person, e.g., how the person behaved in the past (h-individual-probability). While a 2-euro coin can easily be replaced by another 2-euro coin, people are seen as unique. When it comes to humans, there is a moral difference between making inferences about a person using data from other people in support of a general rule and using data only from that person. This moral difference makes h-group probabilities and practices pro tanto morally problematic. If so, using them could require a moral justification, which is not required in a parallel case involving only h-individual probabilities and practices. Typically, this justification will be consequentialist and consider aggregate benefits, disregarding inequalities between groups. As we discussed, the fairness constraints proposed in the literature might provide a further constraint for the moral acceptability of practices that are pro tanto wrong, when a consequentialist argument favors them. This further constraint may be interpreted as a requirement of balancing the harm of misclassification across the different groups.

## AUTHORS CONTRIBUTION STATEMENT

All authors contributed to the introduction and the conclusion, Hertweck wrote sections 2 and 5, Heitz wrote section 4, Loi wrote sections 3 and 7, and Viganò wrote sections 6 and 7. Loi coordinated the writing of the first draft and Viganò coordinated the writing of all subsequent drafts and its review and finalized the manuscript. All authors revised and approved the final version.

## ACKNOWLEDGMENTS

This work was supported by the National Research Programme "Digital Transformation" (NRP 77) of the Swiss National Science Foundation (SNSF), grant number 187473 and the European Union's Horizon 2020 research and innovation programme under the Marie Sklodowska-Curie grant agreement No 898322.

## REFERENCES
[1] Brian Hedden. 2021. On statistical criteria of algorithmic fairness. Philosophy & Public Affairs 49, 2 (2021), 209–231. DOI: https://doi.org/10.1111/papa.12189
[2] Moritz Hardt, Eric Price, Eric Price, and Nati Srebro. 2016. Equality of Opportunity in Supervised Learning. In Advances in Neural Information Processing Systems, Curran Associates, Inc. Retrieved from https://proceedings.neurips.cc/paper/2016/file/9d2682367c3935defcb1f9e247a97c0d-Paper.pdf
[3] Barocas, Solon, Hardt, Moritz, and Narayanan, Arvind. Fairness in Machine Learning. Limitations and Opportunities. Incomplete work in progress.
[4] Alexandra Chouldechova. 2017. Fair Prediction with Disparate Impact: A Study of Bias in Recidivism Prediction Instruments. Big Data 5, 2 (2017), 153–163. DOI: https://doi.org/10.1089/big.2016.0047
[5] Jon Kleinberg, Sendhil Mullainathan, and Manish Raghavan. 2016. Inherent Trade-Offs in the Fair Determination of Risk Scores.
[6] Ellery Eells. 1991. Probabilistic Causality. Cambridge University Press.
[7] Reuben Binns, Max Van Kleek, Michael Veale, Ulrik Lyngs, Jun Zhao, and Nigel Shadbolt. 2018. "It's reducing a human being to a percentage"; perceptions of justice in algorithmic decisions. In Conference on Human Factors in Computing Systems - Proceedings, Association for Computing Machinery. DOI: https://doi.org/10.1145/3173574.3173951
[8] Kasper Lippert-Rasmussen. 2011. "We are all Different": Statistical Discrimination and the Right to be Treated as an Individual. Journal of Ethics 15, 1–2 (June 2011), 47–59. DOI: https://doi.org/10.1007/s10892-010-9095-6
[9] Benjamin Eidelson. 2013. Treating People as Individuals. In Philosophical Foundations of Discrimination Law, Deborah Hellman and Sophia Moreau (eds.). Oxford University Press, Oxford.
[10] Sarah Moss. 2018. Probabilistic knowledge. Oxford University Press. DOI: https://doi.org/10.1093/oso/9780198792154.001.0001

---

[10] This is a possible way of reformulating Castro's view [26] about this problem that makes it not only fully compatible with this view, but also its direct ancestor. Castro argues that "if x's treatment of y exposes y to a risk of being misclassified that y could reasonably reject, then x fails to respect y's individuality" [26, p. 423]. It seems that the substance of Castro's argument is compatible with an alternative conclusion, namely: if x's treatment of y exposes y to a risk of being misclassified that y could not reasonably reject, then, *even if* x fails to respect y's individuality, the treatment is *fair* to x. The fulfilment of a fairness constraint $f_n$ would then specify the misclassification risks that all the different groups involved cannot reasonably reject.

[11] As employing h-group probabilities inevitably violates the individual's right to be treated as individuals, group fairness can also be seen as the guarantee that the harms and benefits resulting from violating that right are distributed proportionally across the affected groups.

[11] Adrian A. S. Zuckerman. 1986. Law, Fact Or Justice. Boston University Law Review 66, (1986). Retrieved from https://heinonline.org/HOL/Page?handle=hein.journals/bulr66&id=493&div=&collection=
[12] David T. Wasserman. 1991. The Morality of Statistical Proof and the Risk of Mistaken Liability. Cardozo Law Review 13, (1991). Retrieved from https://heinonline.org/HOL/Page?handle=hein.journals/cdozo13&id=957&div=&collection=
[13] Amit Pundik. 2016. Freedom and Generalisation. Oxford Journal of Legal Studies 37, 1 (July 2016), gqw016. DOI:https://doi.org/10.1093/ojls/gqw016 [14] Federico Picinali. 2016. Base-rates of Negative Traits: Instructions for Use in Criminal Trials. Journal of Applied Philosophy 33, 1 (February 2016), 69–87. DOI: https://doi.org/10.1111/japp.12109
[14] Federico Picinali. 2016. Base-rates of Negative Traits: Instructions for Use in Criminal Trials. Journal of Applied Philosophy 33, 1 (2016), 69–87. DOI: 10.1111/japp.12109.
[15] Mike Redmayne. 2008. Exploring the proof paradoxes. Legal Theory 14, 4 (2008), 281–309. DOI:https://doi.org/10.1017/S1352325208080117
[16] Kaspar Lippert-Rasmussen. 2014. Born Free and Equal? A Philosophical Inquiry into the Nature of Discrimination. Oxford University Press, Oxford. DOI: https://doi.org/10.1093/analys/anv063
[17] Lawrence Blum. 2004. Stereotypes and stereotyping: A moral analysis. Philosophical Papers 33, 3 (2004), 251–289. DOI: https://doi.org/10.1080/05568640409485143
[18] Erin Beeghly. 2018. Failing to Treat Persons as Individuals. Ergo, an Open Access Journal of Philosophy 5, 26 (July 2018), 687–711. DOI: https://doi.org/10.3998/ergo.12405314.0005.026
[19] Rima Basu. 2019. What We Epistemically Owe To Each Other. Philosophical Studies 176, 4 (2019), 915–931. DOI: https://doi.org/10.1007/s11098-018-1219-z
[20] Richard J. Arneson. 2006. What Is Wrongful Discrimination. San Diego Law Review 43, (2006). Retrieved from https://heinonline.org/HOL/Page?handle=hein.journals/sanlr43&id=799&div=&collection=
[21] Michael Levin. 1992. Responses to Race Differences in Crime. Journal of Social Philosophy 23, 1 (March 1992), 5–29. DOI:https://doi.org/10.1111/j.1467-9833.1992.tb00481.x
[22] Frederick F. Schauer. 2003. Profiles, probabilities, and stereotypes. Harvard University Press, Cambridge-London.
[23] David Miller. 1999. Principles of Social Justice. Harvard University Press.
[24] Shelly Kagan. 1988. The Additive Fallacy. Ethics 99, 1 (October 1988), 5–31. DOI: https://doi.org/10.1086/293033
[25] Mathias Risse and Richard Zeckhauser. 2004. Racial Profiling. Philosophy & Public Affairs 32, 2 (2004), 131–170. DOI: https://doi.org/10.1111/j.1088-4963.2004.00009.x
[26] Clinton Castro. 2019. What's Wrong with Machine Bias. Ergo, an Open Access Journal of Philosophy 6, (2019). DOI: http://dx.doi.org/10.3998/ergo.12405314.0006.015